\documentclass[prd,showpacs,11pt,nofootinbib]{revtex4}
\usepackage{graphicx,color}
\begin{document}
\title{\bf\Large Diquarks and the Semi-Leptonic Decay of
$\Lambda_{b}$ in the Hybrid Scheme}

\vspace{1cm}

\author{Peng Guo$^1$, Hong-Wei Ke$^1$, Xue-Qian Li$^1$, Cai-Dian
L\"u$^2$ and Yu-Ming Wang$^{1,2}$ }

\vspace{1cm}

\address{1. Department of Physics, Nankai University, Tianjin, 300071,
P.R. China
\\
2. CCAST (World Laboratory), P.O. Box 8730, Beijing 100080, China
and \\
Institute of High Energy Physics, CAS, P.O. Box 918, Beijing 100049,
P.R. China}

\vspace{1cm}

\date{\today}

\begin{abstract}

In this work we use the heavy-quark-light-diquark picture to study
the semileptonic decay $\Lambda_b \to \Lambda_c+l+\bar{\nu}_l$ in
the so-called hybrid scheme. Namely, we apply the heavy quark
effective theory (HQET) for larger $q^2$ (corresponding to small
recoil), which is the invariant mass square of $l+\bar\nu$, whereas
the perturbative QCD approach for smaller $q^2$ to calculate the
form factors. The turning point where we require the form factors
derived in the two approaches to be connected, is chosen near
$\rho_{cut}=1.1$. It is noted that the kinematic parameter $\rho$
which is usually adopted in the perturbative QCD approach, is in
fact exactly the same as the recoil factor $\omega=v\cdot v'$ used
in HQET where $v$, $v'$ are the four velocities of $\Lambda_b$ and
$\Lambda_c$ respectively. We find that the final result is not much
sensitive to the choice, so that it is relatively reliable.
Moreover, we apply a proper numerical program within a small range
around $\rho_{cut}$ to make the connection sufficiently smooth and
we parameterize the form factor by fitting the curve gained in the
hybrid scheme. The expression and involved parameters can be
compared with the ones gained by fitting the experimental data. In
this scheme the end-point singularities do not appear at all. The
calculated value is satisfactorily consistent with the data which is
recently measured by the DELPHI collaboration within two standard
deviations.

\end{abstract}
 \maketitle
\vspace{1cm}

\section{Introduction}

The general theory of QCD has been developed for more than 40
years, and at present, nobody ever doubts its validity. However,
on the other side there is still not a reliable way to deal with
the long-distance effects of QCD which are responsible for the
quark confinement and hadronic transition matrix elements, because
their evaluations cannot be done in perturbative approach. Thus,
one needs to factorize the perturbative sub-processes and the
non-perturbative parts which correspond to different energy
scales. The perturbative parts are in principle, calculable to any
order within the framework of quantum field theory, whereas the
non-perturbative part must be evaluated by either fitting data
while its universality is assumed, or invoking concrete models.

The perturbative QCD method (PQCD) has been applied to study
processes where transitions from heavy mesons or baryons to light
hadrons are concerned \cite{Sterman,H.N. Li,Li}, namely the PQCD
which includes the Sudakov resummation, is proved to be successful
for handling processes with small 4-momentum transfer $q^2$.
Indeed the processes involving heavy hadrons may provide us with
an opportunity to study strong interaction, because compared to
$\Lambda_{QCD}$ there exist natural energy scales (heavy quark
masses) which can be used to factorize the perturbative
contributions from the non-perturbative effects. On the other
hand, for the processes involving heavy hadrons, at small recoil
region, where $v\cdot v'$ is close to unity ($v$ and $v'$ denote
the four-velocities of the initial and final hadrons), i.e. the
momentum transfer $q^2$ is sufficiently large, the heavy quark
effective theory (HQET) works well due to an extra symmetry
$SU_f(2)\otimes SU_s(2)$ \cite{Isgur}. Therefore the HQET and PQCD
seem to apply at different regions of $q^2$. For a two-body decay,
the momentum transfer is fixed by the kinematics, however, for a
three-body decay, $q^2$ would span the two different regions.

Among all the processes, the semi-leptonic decay of hadrons plays
an important role for probing the underlying principles and
employed models because this process is relatively simple and less
dependent on the non-perturbative QCD effects. Namely leptons do
not participate in strong interaction, and there is no
contamination from the crossed gluon-exchanges between quarks
residing in different hadrons which are produced in the weak
transitions, whereas such effects are important for the
non-leptonic decays. Thus one might gain more model-independent
information, such as extraction of the Cabibbo-Kobayashi-Maskawa
matrix elements from data. In the semi-leptonic decays of heavy
hadrons it is expected to factorize the perturbative and
non-perturbative parts more naturally. Recently the DELPHI
collaboration reported their measurement on the $\Lambda_b$ decay
form factor in the semi-leptonic process $\Lambda_b\rightarrow
\Lambda_c+l+\bar{\nu}_l$ and determined the parameter $\hat\rho^2$
in the Isgur-Wise function $\xi(v\cdot v')=1-\hat\rho^2(1-v\cdot
v')$ \cite{DELPHI}.

There is a flood of papers to discuss the semi-leptonic decays of
heavy mesons and the concerned factorization. By contraries, the
studies on heavy baryons are much behind  \cite{Liu1,Liu2,xhg},
because baryons consist of three constituent quarks and their
inner structures are much more complicated than mesons.  In this
work, we are going to employ the one-heavy-quark-one-light-diquark
picture for the heavy $\Lambda_b$ and $\Lambda_c$ to evaluate the
form factors of this semi-leptonic transition
$\Lambda_b\rightarrow\Lambda_c+l+\bar\nu$. Even though the subject
of diquark is still in dispute, it is commonly believed that the
quark-diquark picture may be a plausible description of baryons
\cite{Wilczek}, especially for the heavy baryons which possess one
or two heavy quarks.

The kinematic region for the semileptonic decay
$\Lambda_b\rightarrow\Lambda_c+l+\bar\nu$ can be characterized by
the quantity $\rho$, which is defined as
\begin{equation}
\rho \equiv \frac{p \cdot p'}{M_{\Lambda_{b}} M_{\Lambda_{c}}},
\end{equation}
where $p,\; p'$  are the four-momenta of $\Lambda_b$ and
$\Lambda_c$ respectively. It is noted that this parameter $\rho$
which is commonly adopted in the PQCD approach is exactly the same
as the recoil factor $\omega=v\cdot v'$ used in the HQET. The
momentum transfer $q^2$ in the process is within the range of
$(m_l+m_{\nu})^2\leq q^2\leq (M_{\Lambda_b}-M_{\Lambda_c})^2$,
equivalently, it is $1\leq \rho\leq
{(M_{\Lambda_b}^2+M_{\Lambda_c}^2)\over
M_{\Lambda_b}M_{\Lambda_c}}\equiv \rho_{max}$. In the framework of
the HQET \cite{Isgur}, this process was investigated by some
authors \cite{Liu1,Liu2,xhg,Holdom}. For larger $q^2$, the HQET
works well, whereas one can expect that for smaller $q^2$, the
PQCD approach applies. In this work, following Ref.\cite{H.N. Li},
we calculate the contribution from the region with small $q^2$,
i.e. $\rho\rightarrow \rho_{max}$, to the amplitude in PQCD. One
believes that the PQCD makes better sense in this region. K\"orner
et al. discussed similar cases and suggested that the symmetry for
smaller recoil is different from that for larger recoil, so they
used the Isgur-Wise function to obtain the amplitude in the
kinematic region of smaller recoil, but Brodsky-Lepage function
for larger recoils \cite{Kroll}. Our strategy is similar that we
apply the PQCD for small $q^2$ while apply HQET for large $q^2$
where the PQCD is no longer reliable, instead.

Concretely, when integrating the amplitude square from minimum to
maximum of $q^2$ to gain the decay rate, we divide the whole
kinematic region into two parts, small and large $q^2$ ($\rho$,
equivalently). We phenomenologically adopt a turning point at a
certain $\rho$ value, to derive the form factors (defined below in
the text) in terms of PQCD in the region from $\rho_{max}$ to this
point and then beyond it we use the HQET instead \cite{xhg}. We let
the two parts connect at the turning point. From Ref.~\cite{Li}, we
notice that as $\rho\leq 1.1$, the PQCD result is not reliable, so
that we choose the turning point at vicinity of $\rho=1.1$. To
testify if the choice is reasonable, we slightly vary the values of
the turning point as choosing $\rho=1.05$, $1.10$ and $1.15$ to see
how sensitive the result is to the choice.  Moreover, it is noted
that as $\rho_{cut}=1.1$ and 1.15 are chosen, the two parts connect
almost smoothly. Even though, to make more sense, we adopt a proper
numerical program to make the connection sufficiently smooth, namely
we let not only the two parts connect, but also the derivatives from
two sides are exactly equal. In fact, small differences in the
derivatives are easily smeared out by the program. Later in the
text, we will explicitly show that the final result is not much
sensitive to it, thus one can trust its validity. We name the scheme
as the ``hybrid'' approach. We also parameterize the form factor
with respect to $\rho$ based on our numerical results. In fact, when
we integrate over the whole kinematic range of $\rho$, we just use
the parameterized expression.

Moreover, we not only reevaluate the form factors $f_{1}$ and
$g_{1}$ of the exclusive process in the diquark picture, but also
calculate the form factors $f_{2}$ and $g_{2}$ which were neglected
in previous works \cite{Li}.  So far, the data on the $\Lambda_b$
semi-leptonic decay are only provided by the DELPHI collaboration
\cite{DELPHI} and not rich enough to single out the contributions
from $f_{2}$ and $g_{2}$. More accurate measurement in the future
may offer information about them. Our treatment has another
advantage. In the pure PQCD approach, there is an end-point
divergence at $\rho\to 1$, even though it is mild and the decay rate
which includes an integration over the phase space of final states,
i.e. over $\rho$, is finite. As calculating the contribution from
the region with large $q^2$ ($\rho\to 1)$\footnote{If $m_l$ is not
zero, $\rho$ cannot be exactly 1, thus the superficial singularity
does not exist at all, but the form factors are obviously
proportional to $1/m_l$ which has the singular property.} to the
form factors in terms of the HQET, the end-point divergence does not
exist at all.

We organize our paper as follows, in section II, we derive the
factorization formula for  $\Lambda_{b} \rightarrow\Lambda_{c} l
\bar{\nu}$. Our numerical results are presented in Section III.
Finally, Section IV is devoted to some discussions and our
conclusion.

\section{Formulations}

The amplitude of $\Lambda_{b} \rightarrow \Lambda_{c} l
\bar\nu$decay process is written as:
\begin{equation}
\mathcal{M}=\frac{G_{F}}{\sqrt{2}} V_{cb} \overline{l}
\gamma^{\mu} (1-\gamma_{5}) \nu_{l} < \Lambda_{c}(p') \mid
\overline{c} \gamma_{\mu} (1-\gamma_{5}) b \mid \Lambda_{b}(p) >,
\end{equation}
where $p$ and $p'$ are the momenta of $\Lambda_{b}$ and
$\Lambda_{c}$  respectively. According to its Lorentz structure,
the hadronic transition matrix element can be parameterized as
\begin{eqnarray}\label{s1}
\mathcal{M}_{\mu} & \equiv &  < \Lambda_{c}(p') \mid \overline{c}
\gamma_{\mu} (1-\gamma_{5}) b \mid \Lambda_{b}(p) >   \\ \nonumber
 &=&  \overline{\Lambda}_{c}(p')  \left[ \gamma_{\mu} \left(f_{1}(q^{2})
 + \gamma_{5} g_{1}(q^{2})  \right)+ \sigma_{\mu \nu} \frac{ q^{\nu}}
 {M_{\Lambda_{b}}} \left(f_{2}(q^{2}) + \gamma_{5} g_{2}(q^{2})  \right)
 +\frac{q_{\mu}}{M_{\Lambda_{b}}} (f_{3}(q^{2}) + \gamma_{5}  g_{3}(q^{2})
 )\right] \Lambda_{b}(p),
\end{eqnarray}
where  $q \equiv p-p'$ and $\sigma_{\mu \nu} \equiv
 {[\gamma_{\mu},\gamma_{\nu}]}/{2}$.

For the convenience of comparing with the works in literature, we
rewrite the above equation in the following form according to
Ref.\cite{xhg}
\begin{eqnarray}
\mathcal{M}_{\mu} &=&  \overline{\Lambda}_{c}(p')  \left[
\gamma_{\mu} \left(F_{1}(q^{2})
 + \gamma_{5} G_{1}(q^{2})  \right)+\frac{p_{\mu}}{M_{\Lambda_{b}}} (F_{2}(q^{2})
  + \gamma_{5}  G_{2}(q^{2})
 )\right.\nonumber \\
 &&\left.+ \frac{p'_{\mu}}{M_{\Lambda_{c}}} (F_{3}(q^{2}) + \gamma_{5}  G_{3}(q^{2})
 )
 \right] \Lambda_{b}(p).\label{s2}
\end{eqnarray}

For the case of massless leptons,
\begin{equation}
q_{\mu} \overline{l} \gamma^{\mu} (1-\gamma_{5}) \nu_{l}=0,
\end{equation}
thus the form factors $f_{3}$ and $g_{3}$ result in null
contributions. The contributions from $f_{2}$ and $g_{2}$ were
neglected in previous literature  \cite{Li}, nevertheless in our
work, we will consider their contributions to the matrix elements
and calculate them in terms of the diquark picture and our hybrid
scheme.

The kinematic variables are defined as follows. In the rest frame
of $\Lambda_{b}$
\begin{equation}
p \equiv (p^{+},p^{-},\mathbf{p}_{T}) =
\left(\frac{M_{\Lambda_{b}}}{\sqrt{2}},\frac{M_{\Lambda_{b}}}{\sqrt{2}},
\mathbf{0}_{T}\right),
\end{equation}
and
\begin{equation}
p'  = \left(\frac{\rho +\sqrt{\rho
^{2}-1}}{\sqrt{2}}M_{\Lambda_{c}},\frac{\rho -\sqrt{\rho
^{2}-1}}{\sqrt{2}}M_{\Lambda_{c}},\mathbf{0}_{T}\right),
\end{equation}
the diquark momenta inside  $\Lambda_{b}$ and $\Lambda_{c}$ are
parameterized respectively as
\begin{equation}
k_{1}  = \left(0, \frac{M_{\Lambda_{b}} }{ \sqrt{2}}
x_{1},\mathbf{k_{1}}_{T}\right), \ \ \  k_{2}  = \left( \frac{
M_{\Lambda_{c}}}{\sqrt{2}} \xi_1
x_{2},0,\mathbf{k_{2}}_{T}\right),
\end{equation}
where $\rho \equiv \frac{p \cdot p'}{M_{\Lambda_{b}}
M_{\Lambda_{c}}}$, $\xi_1 \equiv \rho +\sqrt{\rho ^{2}-1}$, $x_{1}
\equiv k_{1}^{-} / p^{-}$ and $x_{2} \equiv k_{2}^{+} / p'^{+}$.
According to the factorization theorem  \cite{Sterman,H.N.
Li,Li,Hsiang-nan Li,Hoi-Lai Yu}, the hadronic matrix element is
factorized in the b-space as
\begin{eqnarray}\label{s3}
\mathcal{M}_{\mu} &=& \int^{1}_{0} d x_{1} d x_{2} \int \frac{
d^{2} \mathbf{b}_{1}}{(2 \pi)^{2}} \frac{d^{2} \mathbf{b}_{2}}{(2
\pi)^{2}}
\overline{\Psi}_{\Lambda_{c}}(x_{2},{\mathbf{b}_{2}},p',\mu)
\nonumber \\
&& \times
\widetilde{H}_{\mu}(x_1,x_2,{\mathbf{b}_{1}},{\mathbf{b}_{2}},r,M_{\Lambda_b},\mu)
\Psi_{\Lambda_{b}}(x_{1},{\mathbf{b}_{1}},p,\mu),
\end{eqnarray}
where $r \equiv M_{\Lambda_{c}} / M_{\Lambda_{b}}$. The
renormalization group evolution of the hard amplitude
$\widetilde{H}_{\mu}$ is shown as follows \cite{Hsiang-nan Li
sudakov}
\begin{eqnarray}
&&\widetilde{H}_{\mu}(x_1,x_2,{\mathbf{b}_{1}},{\mathbf{b}_{2}},r,M_{\Lambda_b},\mu)
={\rm{exp}}[-4 \int_{\mu}^{t_{a(b)}} \frac{d
\overline{\mu}}{\overline{\mu}}\gamma_q(\alpha_{s}(\overline{\mu}))]
\widetilde{H}_{\mu}(x_1,x_2,{\mathbf{b}_{1}},{\mathbf{b}_{2}},r,M_{\Lambda_b},t_{a(b)}),
\end{eqnarray}
where $\gamma_q(\alpha_{s}(\overline{\mu}))$ is the anomalous
dimension.

The wave function of $\Lambda_{b}$ which has the heavy-quark and
light-diquark structure, is given as \cite{Kroll,kroll}
\begin{eqnarray}\label{f1}
 \Psi_{\Lambda_{b}}(x_{1},\mathbf{b}_{1},p,\mu ) &=&
  f^{S}_{\Lambda_{b}} \Phi^{S}_{\Lambda_{b}} (x_{1},\mathbf{b}_{1},p,\mu)
 \chi^{S}_{\Lambda_{b}} \Lambda_{b}(p,\lambda) ,
\end{eqnarray}
where $\Lambda_{b}(p,\lambda)$ is  the baryon spinor, and the
superscript S  denotes  scalar diquark (spin$=0$, isospin$=0$).
$f^{S}_{\Lambda_{b}}$ is a constant introduced in literature.
$\chi^{S}_{\Lambda_{b}}$ is the flavor component of the baryon,
namely $\chi^{S}_{\Lambda_{b}}=b^{\dagger} S^{\dagger}_{[u,d]}|0>$,
where $b^{\dagger}$ and $S^{\dagger}_{[u,d]}$ are the creation
operator of b-quark and the scalar diquark of $ud-$quarks.

The $\Lambda_{c}$ distribution amplitude bears a similar form,
\begin{eqnarray}\label{f2}
 \Psi_{\Lambda_{c}}(x_{2},{\mathbf{b}_{2}},p',\mu) &=&
  f^{S}_{\Lambda_{c}} \Phi^{S}_{\Lambda_{c}} (x_{2},{\mathbf{b}_{2}},p',\mu)
 \chi^{S}_{\Lambda_{c}} \Lambda_{c}(p',\lambda_2) .
\end{eqnarray}
Including the Sudakov evolution of hadronic wave functions, i.e.
running the scale of wave function from $\mu$ down to $\omega_1
(\omega_2)$ \cite{Hsiang-nan Li sudakov}:
\begin{eqnarray}
&&\Phi^{S}_{\Lambda_b}(x_{1},{\mathbf{b}_{1}},p,\mu)=\mathrm{exp}[-s(\omega_1,(1-x_l)
p^-)-2\int_{\omega_1}^{\mu} \frac{d
\overline{\mu}}{\overline{\mu}}\gamma_q(\alpha_{s}(\overline{\mu}))
]\Phi^{S}_{\Lambda_b}(x_{1},{\bf{b}_1}), \nonumber \\
&&\Phi^{S}_{\Lambda_c}(x_2,{\mathbf{b}_{2}},p',\mu)=\mathrm{exp}[-s(\omega_2,(1-x_2)
p'^+)-2\int_{\omega_2}^{\mu} \frac{d
\overline{\mu}}{\overline{\mu}}\gamma_q(\alpha_{s}(\overline{\mu}))
]\Phi^{S}_{\Lambda_c}(x_{2},{\bf{b}_2})\label{suda},
\end{eqnarray}
where $\omega_{i}=1/b_{i}$(i=1,2).

 In our work,
the effective gluon-diquark vertices are defined as
 \cite{Kroll,kroll}
\begin{eqnarray}
 SgS:&& i g_{s} t^{\alpha} (p_{1}+p_{2})_{\mu} F_{S}(Q^{2}),\nonumber  \\
 && F_{S}(Q^{2})= \delta_s \frac{ Q_{0}^{2} }{ Q_{0}^{2}+Q^{2} }, \nonumber \\
&& \delta_s = \alpha_{s}(Q^{2})/  \alpha_{s}(Q_{0}^{2})\ \ (if \ \
Q^{2} \geq Q_{0}^{2});\qquad \delta_s = 1 \ \ (if \ \ Q^{2} <
Q_{0}^{2}),
\end{eqnarray}
where $Q^{2} \equiv -(p_{1}-p_{2})^{2}$.

\begin{figure}[t]
\begin{center}
\begin{tabular}{ccc}
\scalebox{1.0}{\includegraphics{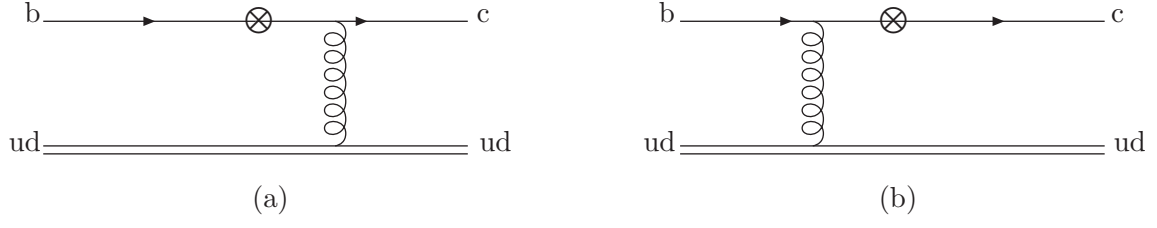}}
\end{tabular}
\end{center}
\caption{The lowest order diagrams of the hard parts of the
transition processes in the quark-diquark picture.}\label{t1}
\end{figure}

According to the factorization scheme which is depicted in
Fig.\ref{t1}, it is straightforward  to obtain the analytic
expressions of the form factors $F_{1}$, $G_{1}$, $F_{2}$, $G_{2}$,
$F_{3}$ and $G_{3}$, by  comparing  eqs. (\ref{s2}) with (\ref{s3}).
In the above derivation, the following transformation has been used.
\begin{eqnarray}
k_1=A p+ B p', \qquad  k_2=C p+ D p'
\end{eqnarray}
and the explicit expressions of A,B,C and D are given in the
appendix.

Then we can obtain the analytical form of $f_{i}$ and  $g_{i}$
(i=1,2,3) making use of the relations between them and $F_{j},
G_{j}$(j=1,2,3) listed below
\begin{eqnarray}\label{gx}
f_1&=&F_1+\frac{1}{2}(\frac{F_2}{M_{\Lambda_{b}}}+\frac{F_3}{M_{\Lambda_{c}}})
(M_{\Lambda_{b}}+M_{\Lambda_{c}}),\ \ g_1=G_1-\frac{1}{2}(\frac{G_2}
{M_{\Lambda_{b}}}+\frac{G_3}{M_{\Lambda_{c}}})(M_{\Lambda_{b}}-M_{\Lambda_{c}}),\nonumber\\
f_2&=&-\frac{1}{2}(\frac{F_2}{M_{\Lambda_{b}}}+\frac{F_3}{M_{\Lambda_{c}}})M_{\Lambda_{b}}^2,
\ \ g_2=-\frac{1}{2}(\frac{G_2}{M_{\Lambda_{b}}}+\frac{G_3}{M_{\Lambda_{c}}})M_{\Lambda_{b}}^2,\nonumber\\
f_3&=&\frac{1}{2}(\frac{F_2}{M_{\Lambda_{b}}}-\frac{F_3}{M_{\Lambda_{c}}})M_{\Lambda_{b}}^2,\
\
g_3=\frac{1}{2}(\frac{G_2}{M_{\Lambda_{b}}}-\frac{G_3}{M_{\Lambda_{c}}})M_{\Lambda_{b}}^2.
\end{eqnarray}
We do not display the expressions of $f_{3}$ and $g_{3}$ for the
reason given above. The form factors are integrations which
convolute over three parts \cite{Hsiang-nan Li}: the hard-part
kernel function, the Sudakov factor and the wave functions of the
concerned hadrons as
\begin{eqnarray}
 f_{i}(g_{i})&=& 4 \pi C_{F}  f_{\Lambda_{b}}f_{\Lambda_{c}}
  \int^{1}_{0} dx_{1} dx_{2} \int^{\infty}_{0} b_{1}
 db_{1}
 b_{2} db_{2} {\int_0}^{2\pi}d\theta\ \Phi_{\Lambda_{c}}(x_{2},\mathbf{b}_{2}) \\
 && \times kernSgS_{f_{i}(g_{i})} \Phi_{\Lambda_{b}}(x_{1},\mathbf{b}_{1})
 \exp {[-S(x_1,x_2,{\mathbf{b}_{1}},{\mathbf{b}_{2}},r,M_{\Lambda_b})]}, \nonumber
\end{eqnarray}
where the explicit expression of the kernel functions
$kernSgS_{f_{i}(g_{i})}$ is given in the appendix for concision of
the text. The explicit form of the Sudakov factor appearing in the
above equations is given in Ref.\cite{Hsiang-nan Li} as
\begin{eqnarray}
&&S(x_1,x_2,{\mathbf{b}_{1}},{\mathbf{b}_{2}},r,M_{\Lambda_b})\nonumber \\
&&=s(\omega_1,(1-x_l) p^-)+s(\omega_2,(1-x_2)
p'^+)+2\int_{\omega_1}^{t_{a(b)}} \frac{d
\overline{\mu}}{\overline{\mu}}\gamma_q(\alpha_{s}(\overline{\mu}))+2\int_{\omega_2}^{t_{a(b)}}
\frac{d
\overline{\mu}}{\overline{\mu}}\gamma_q(\alpha_{s}(\overline{\mu})),
\end{eqnarray}
where $C_{F}=4/3$ is the color factor.

The wave function $\Phi_{\Lambda_{b}}(x_{1},\mathbf{k}_{1T})$ is
\cite{Hoi-Lai Yu,peterson}
\begin{eqnarray}
\Phi_{\Lambda_{b}}(x_{1},\mathbf{k}_{1T})= \frac{2 N_{b}
x_{1}^{6}(1-x_{1})^{3}}{\pi [(1-a_{b}-x_{1})^{2}x_{1}^{2}(1-x_{1})
+ \varepsilon_{pb} (1-x_{1})^{2}x_{1}^{2}+
\mathbf{k}_{1T}^{2}]^{3}},
\end{eqnarray}
and
\begin{eqnarray}
\Phi_{\Lambda_{b}}(x_{1},b_{1})&=& \int d^{2}
\mathbf{k}_{1T}\Phi_{\Lambda_{b}}(x_{1},\mathbf{k}_{1T}) e^{i
\mathbf{k}_{1T} \cdot \mathbf{b}_{1} } \nonumber \\
&=&\frac{ N_{b} x_{1}^{6}(1-x_{1})^{3} b_{1}^{2}
K_{2}(\sqrt{(1-a_{b}-x_{1})^{2}x_{1}^{2}(1-x_{1}) +
\varepsilon_{pb} (1-x_{1})^{2}x_{1}^{2}}b_{1})}{2
[(1-a_{b}-x_{1})^{2}x_{1}^{2}(1-x_{1}) + \varepsilon_{pb}
(1-x_{1})^{2}x_{1}^{2}]},
\end{eqnarray}
where $K_{2}$ is the modified Bessel function of the second kind.
If neglecting the transverse momentum  $\mathbf{k}_{1T}$, i.e. set
$\mathbf{k}_{1T} \sim 0$, the wave function can be simplified as
\begin{eqnarray}
\Phi_{\Lambda_{b}}(x_{1}) \simeq
\frac{N_{b}x_{1}^{2}(1-x_{1})}{[(1-a_{b}-x_{1})^{2} +
\varepsilon_{pb} (1-x_{1})]^{2}}.
\end{eqnarray}
The normalization conditions are set as  \cite{Hoi-Lai Yu}
\begin{eqnarray}
\int_{0}^{1} \Phi_{\Lambda_{b}}(x_{1})   dx_{1}
&=&1 , \nonumber \\
\int_{0}^{1} \Phi_{\Lambda_{b}}(x_{1}) x_{1}  dx_{1}
&=&\frac{\bar{\Lambda}}{M_{\Lambda_{b}}} , \nonumber \\
\int_{0}^{1} \Phi_{\Lambda_{b}}(x_{1}) x_{1}^{2}  dx_{1}
&=&(\frac{\bar{\Lambda}^{2}}{M_{\Lambda_{b}}^{2}}+
\frac{\lambda_{1}}{3 M^2_{\Lambda_{b}}} ) .
\end{eqnarray}
The first formula determines the normalization of the parton
distribution of the baryon, whereas the second one is related to
the effective mass of the light diquark $\bar{\Lambda} \sim
M_{\Lambda_{b}} -m_{b} $, and the third formula reflects
connection between hadronic matrix element of the kinematic
operator $\lambda_{1}=-\frac{1}{2
M_{\Lambda_{b}}}\langle\Lambda_{b}\mid \bar{b}_{v} (i
D_{\perp})^{2} b_{v} \mid \Lambda_{b}\rangle $  and hadronic
distribution amplitude. To satisfy the above three normalization
conditions, the parameters would take the following values
$\bar{\Lambda}=0.848$ GeV, $\lambda_{1}=-0.76$ GeV$^2$,
$a_{b}=0.916$, $\varepsilon_{pb}=0.0051$ and $N_{b}=0.0219$, and
in the following numerical evaluation, we will use them as inputs.

For the  wave function of $\Lambda_{c}$, the expressions are the
same as that for $\Lambda_{b}$, while the corresponding parameters
are $\bar{\Lambda}=0.8849$ GeV, $\lambda_{1}=-1.87 $GeV$^2$ ,
$a_{c}=1.48$, $\varepsilon_{pc}=0.080$ and $N_{c}=12.14$. Thus we
can write the differential decay width as
\begin{eqnarray}
\frac{d \Gamma }{ d \rho} &=&\frac{G_{F}^{2} | V_{cb}|^{2}}{24
\pi^{3}} M_{\Lambda_{b}}^{5} r^{3} \sqrt{\rho^{2}-1}\nonumber \\
&& [ \ |f_{1}|^{2} (\rho-1)(3 r^{2}-4 \rho r+2 r +3)  +
|g_{1}|^{2}
(\rho+1)(3 r^{2}-4 \rho r-2 r +3)  \nonumber \\
&&  +6\ f_1\ f_2(r+1)(\rho-1)(r^2-2\ r\rho+1)+6\ g_1\ g_2(r-1)(r^2-2\ r\rho+1)\nonumber \\
&& + |f_{2}|^{2} (\rho-1)(3 r^{2}-2 \rho r+4 r +3)(r^{2}-2 \rho r
+1)\nonumber \\
&& + |g_{2}|^{2} (\rho+1)(3 r^{2}-2 \rho r-4 r +3)(r^{2}-2 \rho r
+1) \ ],
\end{eqnarray}
 with
\begin{equation}
 1 \leq \rho \leq \rho_{max}=
 \frac{1}{2}\left(\frac{M_{\Lambda_{b}}}{M_{\Lambda_{c}}}+\frac{M_{\Lambda_{c}}}{M_{\Lambda_{b}}}\right).
\end{equation}

\section{Numerical Results}

\subsection{The results of PQCD}

\begin{figure}[t]
\begin{center}
\scalebox{0.8}{\includegraphics{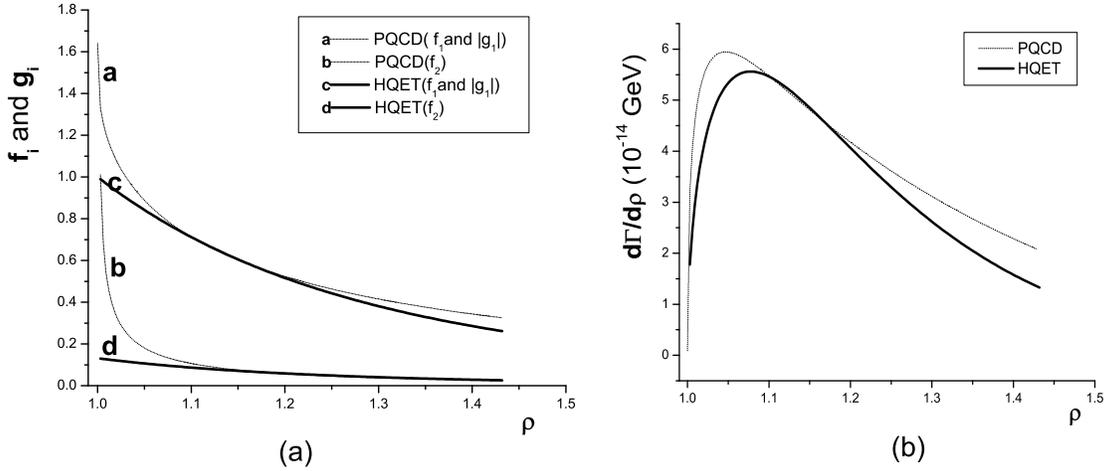}} \vspace{-1cm}
\end{center}
\caption{(a) Form factors $f_2$, $ f_1$ and $|g_1|$ \,\,\,  (b)
Differential decay width of $\Lambda_{b} \rightarrow \Lambda_{c} l
\nu$}\label{t2}
\end{figure}

In the one-heavy-quark-one-light-diquark picture, the $(ud)_{\bar
3}$ diquark in $\Lambda_{b}$ is considered as a scalar of color
anti-triplet. To calculate the form factors in the framework of
PQCD, one can adjust the product
$f^{S}_{\Lambda_{b}}f^{S}_{\Lambda_{c}}$ to fit the empirical
formula $f_{1}(\rho_{max}) \sim 1.32/\rho_{max}^{5.18}$ given by the
authors of Ref.\cite{Li}\footnote{When first calculating the form
factors, there were no data available, the authors of \cite{Li} used
a reasonable estimate of $BR(\Lambda_{b} \rightarrow \Lambda_{c} l
\nu)$ as about $2\%$. Nowadays, measurements have been done and with
the data, we have made a new fit. }. Then the corresponding
parameters are obtained as
\begin{eqnarray}\label{PQCD}
   m_{b} \simeq M_{\Lambda_{b}}=5.624\;{\rm GeV}, &m_{c} \simeq
 M_{\Lambda_{c}}=2.2849\;{\rm GeV}, & V_{cb}=0.040,\nonumber
 \\
 f^{S}_{\Lambda_{b}}f^{S}_{\Lambda_{c}} =0.0096 \rm{GeV^2},&
   Q^{2}_{0} =3.22 \rm{GeV^{2}},&
 \Lambda_{QCD}=0.2 \rm{GeV}.
\end{eqnarray}
Using these values, we can continue to numerically estimate the form
factors $f_1$ and $g_1$. Fig.\ref{t2}(a) shows that the form factor
$f_1$ is exactly equal to $|g_1|$ in the heavy quark limit. The form
factor $f_2$ and $g_2$ are much smaller than $f_1$ and $|g_1|$, thus
they can in fact be safely neglected for the present experimental
accuracy.

In  Fig.\ref{t2}(b), we plot the dependence of the differential
width $d\Gamma/d\rho$ on $\rho$. Although both $f_1$ and $g_1$ have
an end-point divergence at $\rho\rightarrow 1$ in PQCD approach, the
differential decay rate is finite.

If one extrapolates the PQCD calculation to the region with
smaller $\rho$ values, we obtain the form factors $f_{1,2}$ and
$g_{1,2}$ in that region where there are obvious end-point
singularities at $\rho\to 1$. Redo the computations with the
extrapolation (in the original work \cite{Li}, the authors extend
the tangent of the PQCD result at a small $\rho$ value to
$\rho=1$, so the end-point singularity is avoided) and obtain
$BR(\Lambda_{b} \rightarrow \Lambda_{c} l \nu)$ which is about
$1.35 \%$ (slightly smaller than the value of 2\% guessed by the
authors of Ref.\cite{Li}, because then no data were available. ).

Obviously, the calculation in PQCD depends on the factor
$f^{S}_{\Lambda_{b}}f^{S}_{\Lambda_{c}}$, which regularly must be
obtained by fitting the data of semileptonic decays, so that the
theoretical predictions are less meaningful. Instead, we will use
our hybrid scheme where we do not need to obtain the factor
$f^{S}_{\Lambda_{b}}f^{S}_{\Lambda_{c}}$ by fitting data, since
the connection requirement substitutes the fitting procedure (see
below for details).

\subsection{The results of HQET}

The transition rate was evaluated in terms of HQET  by the authors
of Ref.~\cite{xhg}. According to the definitions given in
eq.(\ref{gx}), we re-calculate $f_1,\; f_2,\; g_1,\; g_2$ while
dropping out $f_3$ and $g_3$ and also obtain similar conclusion that
$f_1$ and $g_1$ are the same in amplitude, but opposite in sign, as
shown in Fig. \ref{t2}(a), whereas $f_2$ is very small and $g_2$ is
exactly zero in HQET. The theoretical prediction on the rate of the
semi-leptonic decay $\Lambda_b\rightarrow\Lambda_c+l+\bar\nu$ in the
HQET is

\begin{eqnarray}
 \Gamma(\Lambda_{b} \rightarrow \Lambda_{c} l \bar\nu)&=& 1.54 \times
 10^{-14}\; {\rm{GeV}},\;\; {\rm without\; contributions\; from}\; f_2 \; {\rm and}\; g_2,
\nonumber
 \\
\Gamma(\Lambda_{b} \rightarrow \Lambda_{c} l \bar\nu) &=&
1.56\times
 10^{-14}
 \;{\rm{GeV}},\;\; {\rm with\; contributions\; from}\; f_2 \; {\rm and}\; g_2,
\end{eqnarray}
and the branching ratio is $2.87 \%$.

The transition rate of $\Lambda_{b} \rightarrow \Lambda_{c} l
\bar\nu$ has recently been measured by the DELPHI Collaboration
\cite{DELPHI}, and the  value of $BR(\Lambda_{b} \rightarrow
\Lambda_{c} l \nu)$ is
$5.0^{+1.1}_{-0.8}(stat)^{+1.6}_{-1.2}(syst) \%$. It is noted that
the result calculated in terms of HQET is only consistent with
data within two standard deviations.

\subsection{The hybrid scheme: Reconciling the two approaches}

\begin{figure}[t]
\begin{center}
\scalebox{0.8}{\includegraphics{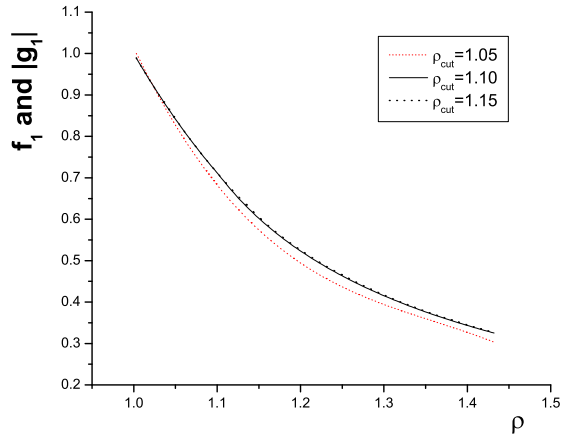}} \vspace{-1cm}
\end{center}
\caption{Form factors  $ f_1=|g_1|$ in the hybrid scheme }\label{t3}
\end{figure}

As widely discussed in literature, in the region with large $q^2$
(small $\rho-$values), the result of HQET is reliable, whereas for
the region with small $q^2$ (larger $\rho-$values) the PQCD is
believed to work well. Therefore, to reconcile the two approaches
which work in different $\rho$ regions, $1\leq\rho<\rho_{max}$, we
apply the HQET for small $\rho$, but use PQCD for larger $\rho$.
Our strategy is that we let the form factors $f_1$, $g_1$ derived
in the PQCD approach be equal to the value obtained in terms of
HQET at the point $\rho_{cut}$. The numerical results of the form
factor $f_1=|g_1|$ in the hybrid scheme are shown in Fig.\ref{t3}
for three different $\rho_{cut}-$values: 1.05, 1.10 and 1.15,
respectively. It is noted that for $\rho_{cut}=1.1$ and 1.15, the
left- and right-derivatives are very close and the connection is
smooth, whereas, as $\rho_{cut}$ is chosen as 1.05, a difference
between the derivatives at the two  sides of $\rho_{cut}=1.05$
obviously manifests. We then adopt a proper numerical program to
smoothen the curve, namely let the derivatives of the two sides
meet each other for any $\rho-$value near the cut point.
Fig.\ref{t3} shows that such treatment in fact does not change the
general form of the curve, but makes it sufficiently smooth for
all $\rho$ values including the selected cut point $\rho_{cut}$.

Another advantage of adopting such a ``hybrid'' scheme is that there
does not exist end-point singularity for the form factors at
$\rho=1$. Since at the turning point, we let the form factors
derived in terms of PQCD be connected with that obtained in HQET,
the product $f_{\Lambda_b}f_{\Lambda_c}$ is automatically determined
by the connection. With the value, we calculate the form factors
within the range of small $q^2$ in PQCD. In this hybrid scheme, one
does not need to invoke the data on the semi-leptonic decay to fix
the parameter $f_{\Lambda_b}f_{\Lambda_c}$ at all.

By our numerical results obtained in the hybrid scheme, the form
factors $f_1$ (or $ g_1$) can be parameterized in a satisfactory
expression, here we only present the expression for
$\rho_{cut}=1.10$  as
\begin{equation}\label{para}
f_1(\rho)=1-3.61(\rho-1)+7.24(\rho-1)^2-5.83(\rho-1)^3,
\end{equation}
and similar parameterized form factors were discussed in
ref.\cite{cheng}.

The expression can be described by only one ``Isger-Wise
function'' for the transition $\Lambda_b\rightarrow \Lambda +l\bar
l$  at the heavy quark limit, and it is done by the DELPHI
collaboration based on their data on $\Lambda_{b} \rightarrow
\Lambda_{c} l \bar\nu$. It is parameterized as \cite{DELPHI}
\begin{equation}\label{para2}
\xi(\omega)=1-\hat\rho^2(\omega-1)+O((\omega-1)^2),
\end{equation}
where $\hat\rho^2=2.03\pm0.46(stat)^{+0.72}_{-1.00}(syst)$.

Obviously, this expression is only valid to the leading order,
i.e. linearly proportional to $\omega-1$ where $\omega$ exactly
corresponds to the parameter $\rho$ which is commonly adopted in
the PQCD language. By contrast, our result includes higher power
terms because the $1/M$ corrections are automatically taken into
account in our work. It is noted that the coefficient of the
linear term in our numerical result is reasonably consistent with
the $\hat\rho^2$ obtained by fitting data.

To obtain the total decay width, we integrate over the whole range
$\rho$ from 1 to $\rho_{max}$, the integrand is the parameterized
form factor in eq.(\ref{para}). We obtain
\begin{equation}
\Gamma(\Lambda_{b} \rightarrow \Lambda_{c} l \nu) = 1.65 \times
10^{-14} {\rm{GeV}},\;\;BR=3.08\%\;\;\;\; \;{\rm with}\;
\rho_{cut}=1.10,\;\;f_{\Lambda_b}f_{\Lambda_c}=0.0149\;
{\rm{GeV}^2}.
\end{equation}
For a comparison, we present the results corresponding to other
two $\rho_{cut}$ values where the smoothing treatment is employed,
and they are
\begin{eqnarray}
\Gamma(\Lambda_{b} \rightarrow \Lambda_{c} l \nu)&=& 1.54 \times
10^{-14} \;{\rm{GeV}},\;\;BR=2.87\%\;\; \;\; \;
\rho_{cut}=1.05,\;\; f_{\Lambda_b}f_{\Lambda_c}=0.0142\;\rm{
{GeV}^2};
\nonumber\\
\Gamma(\Lambda_{b} \rightarrow \Lambda_{c} l \nu) &=& 1.67 \times
10^{-14}\; {\rm{GeV}},\;\; BR=3.12\%\;\;\;\; \;
\rho_{cut}=1.15;\;\;f_{\Lambda_b}f_{\Lambda_c}=0.0150\;
\rm{{GeV}^2}.
\end{eqnarray}

One can notice that the factor $f_{\Lambda_b}f_{\Lambda_c}$ does
not change much as $\rho_{cut}$ varies and they are about 1.5
times larger than the value obtained in pure PQCD
(eq.(\ref{PQCD})).  When the turning point is chosen at
$\rho_{cut}=1.05$, the branching ratio calculated in the hybrid
scheme is very close to the result obtained in pure HQET, while
for $\rho_{cut}=1.10$ and $\rho_{cut}=1.15$, the resultant
branching ratio is slightly larger than that obtained in pure
HQET, but more coincides with the data.

\begin{figure}[t]
\begin{center}
\scalebox{0.8}{\includegraphics{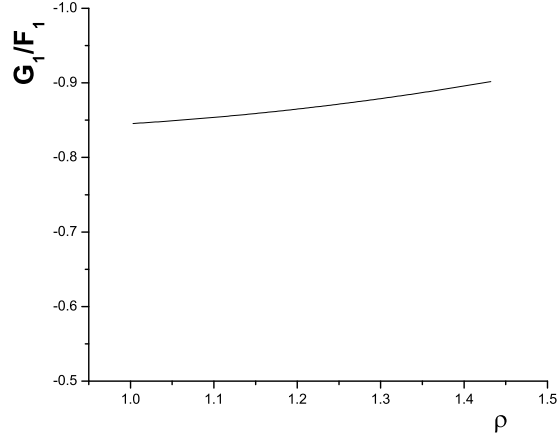}}
\end{center}
\vspace{-1cm} \caption{The $G_1/F_1$ ratio obtained in our hybrid
scheme }\label{t5}
\end{figure}

In our scenario, the HQET is applied for smaller $\rho$, and the
values of the form factors at $\rho_{cut}$ are fixed by the theory.
The values can also determine $f_{\Lambda_b}f_{\Lambda_c}$ which
will be used for the PQCD calculations for larger $\rho$. The HQET
is an ideal theoretical framework, but there is an unknown function
which is fully governed by the non-perturbative QCD effects, that is
the famous Isgur-Wise function. The function can be either obtained
by fitting data, or evaluated by concrete models. Various models
would result in different slopes. The authors of ref. \cite{xhg}
used the Drell-Yan type overlap integrals to obtain the slope which
is what we employed to get the parametrization eq.(\ref{para}) and
the slope is $-3.6$. Instead, the authors of \cite{Liu1} evaluated
the slope in the Isgur-Wise function by means of the QCD sum rules.
According to their result, we re-parameterize the form factor
$f_1(\rho)$ and have
\begin{equation}\label{para1}
f_1(\rho)=1.01-1.57(\rho-1)-2.59(\rho-1)^2+6.99(\rho-1)^3.
\end{equation}
Correspondingly, we obtain
\begin{equation}
\Gamma(\Lambda_{b} \rightarrow \Lambda_{c} l \nu) = 2.38 \times
10^{-14} {\rm{GeV}},\;\;BR=4.45\%\;\;\;\; \;{\rm with}\;
\rho_{cut}=1.10,\;\;f_{\Lambda_b}f_{\Lambda_c}=0.0213\;
{\rm{GeV}^2}.
\end{equation}

The fitted slope by the DELPHI collaboration is
$\hat\rho^2=2.03\pm0.46(stat)^{+0.72}_{-1.00}(syst)$
\cite{DELPHI}, which is between the two theoretical evaluated
values. In these references, only linear term remained, due to
uncertainties in the approximations the deviations are
understandable. Therefore, one can note that there is a
model-dependence which mainly manifests in the slope of the
Isgur-Wise function. Even though they deviate from each other at
the linear term, the high power terms would compensate the
deviation slightly and the predicted values on the branching ratio
in two approaches eqs.(\ref{para},\ref{para1}) are qualitatively
consistent with data. There indeed is a byproduct which brings in
an advantage that more accurate measurements can help to make
judgement on validity of the models by which the Isgur-Wise
functions are evaluated.

To make more sense, we purposely  present the ratio $G_1/F_1$
obtained in the hybrid scheme in Fig. \ref{t5}, it is noted that
the ratio is qualitatively consistent with that given in ref.
\cite{Kroll} which was shown on the left part of Fig. 3 of their
paper \cite{Kroll}.

The authors of ref. \cite{Kroll} extended $\rho$ into the
un-physical region ($\rho>\rho_{max}$ for $\Lambda_{b} \rightarrow
\Lambda_{c} l \bar\nu$), while we only keep it within the physical
region. It is noted that in the physical region, numerically our
result is very close to that obtained in ref. \cite{Kroll}. But if
one extends the curve to larger $\rho$, he will notice that our
curve is convex, but theirs is concave, namely the coefficient of
the quadratic term has an opposite sign, but the difference is too
tiny to be observed or bring up substantial difference for the
evaluation of the decay width.


\section{discussion and conclusion}

In this work, we investigate the semi-leptonic decay $\Lambda_b \to
\Lambda_c+l+\bar{\nu}_l$  in the so called ``hybrid scheme'' and the
diquark picture for heavy baryons $\Lambda_b$ and $\Lambda_c$. The
hybrid scheme means that for the range of smaller $q^2$ (larger
$\rho-$value) we use the PQCD approach, whereas the HQET for larger
$q^2$ (near $\rho=1$), to calculate the form factors. We find that
the form factors $f_2$ and  $g_2$ can be safely neglected as
suggested in the literature. Besides, we do not need  to determine
the phenomenological parameters $f_{\Lambda_b}$ and $f_{\Lambda_c}$
by fitting data in the hybrid scheme as one did with the pure PQCD
approach. Our result is generally consistent with the newly measured
branching ratio $\Lambda_b \to \Lambda_c+l+\bar{\nu}_l$ within two
standard deviations and the end-point singularities existing in the
PQCD approach are completely avoided. In fact, the final result
somehow depends on the slope in the Isgur-Wise function of the HQET,
which is obtained by model-dependent theoretical calculations.
Therefore, we may only trust the obtained value to this accuracy,
the further experimental data and development in theoretical
framework will help to improve the accuracy of the theoretical
predictions.


The quark-diquark picture seems to work well for dealing with the
semi-leptonic decays of $\Lambda_b$, and we may expect that the
quark-diquark picture indeed reflects the physical reality and is
applicable to the processes where baryons are involved, at least
for the heavy baryons  \cite{Guo}. This picture will be further
tested in the non-leptonic decays of heavy baryons. We will employ
the diquark picture and PQCD to further study the non-leptonic
decay modes in our future work.

\vspace{0.8cm}

\section*{Acknowledgement}

This work is partly supported by the National  Science Foundation of
China (NSFC) under contract No.10475042, 10475085 and 10625525. We
thank Dr. C. Liu for helpful discussions.

\appendix

\section{Explicit expressions of hard kernel}
\begin{eqnarray}
 kernSgS_{f_{1}}&=&\Big\{\Big[( B C - A D )( 1 + r )
     -
   ( C^2 - A  (  2 - C + D r  )  +
     (  B + D  ) r
      (  D r - 2 \rho  )  +\nonumber\\&&
     C(  B r + 2B r \rho +
        2D r\rho    -2)  )\Big]hha+\Big[(  B  C - A  D )   r\
     ( 1 +   r )
     - \
   ( A^2 + A  \nonumber\\&& ( C -
        r  ( D + 2  \rho - 2  B  \rho  )  )
     + r   (  B^2  r -
        2  (  D  r + C  \rho  )  -
        B  (  C + ( -2 + D )   r\nonumber\\&& +
           2  C  \rho  )  )  )\Big]hhb\Big\}{M^2_{\Lambda_{b}}}F_{S}(\eta), \nonumber \\
 kernSgS_{g_{1}}&=&\Big\{\Big[(  BC - AD )
     (  r -1)
       +
   ( C^2 + A( -2 + C + Dr )  +
     ( B + D ) r
      ( Dr - 2\rho )  -\nonumber\\&&
     C(  r B - 2Dr\rho - 2 rB\rho  +2 )  )
\Big]hha+\Big[(   B  C   - A  D ) \
      r
     (    1-r )
       +
   ( A^2 + A \nonumber\\&& ( C +
        r  ( D  -2 \rho + 2 B\rho
           )  )  +
     r  ( B^2  r - 2  ( D  r + C  \rho )  +
       B\  ( ( -2 + D )  r
             -C\nonumber\\&&  + 2 C \rho   )  )
     )\Big]hhb\Big\}\ {M^2_{\Lambda_{b}}}F_{S}(\eta), \nonumber \\
  kernSgS_{f_{2}}&=&  ( A D - B C   )\ (
hha+  r\
    hhb )\ M^2_{\Lambda_{b}} F_{S}(\eta), \nonumber \\
  kernSgS_{g_{2}}&=& ( A D  - B C  )\  (
  hha-  r \  hhb) \  M^2_{\Lambda_{b}} F_{S}(\eta), \nonumber \\
\end{eqnarray}

where
\begin{eqnarray}
\eta \equiv M_{\Lambda_{b}}M_{\Lambda_{c}}x_{1}x_{2}\xi_1.
\end{eqnarray}

The explicit expressions of A, B, C and D
\begin{eqnarray}
A=\frac{x_1\ \xi_1}{\xi_1-\xi_2},\qquad \ B=-\frac{x_1\
M_{\Lambda_{b}}}{M_{\Lambda_{c}}(\xi_1-\xi_2)},\nonumber\\C=-\frac{x_2\
M_{\Lambda_{c}}}{M_{\Lambda_{b}}(\xi_2-\xi_1)},\qquad \
D=\frac{x_2\ \beta_1}{\xi_1-\xi_2},
\end{eqnarray}

with
\begin{equation}
\xi_1=\rho+\sqrt{\rho^2-1}, \qquad \xi_2=\rho-\sqrt{\rho^2-1}.\nonumber\\
\end{equation}

The explicit expressions of $hha$, $hhb$ are
\begin{eqnarray}
hha&=&\alpha_{s}(t_a) K_{0} (\sqrt{x_{1} x_{2} \xi_1
M_{\Lambda_{b}} M_{\Lambda_{c}}}b_{1})  K_{0}( \sqrt{ x_{2} \xi_1
M_{\Lambda_{b}}
M_{\Lambda_{c}}}|\textbf{b}_1+\textbf{b}_2|),\,\,\,\,\,\,\nonumber\\
hhb&=&\alpha_{s}(t_b) K_{0} (\sqrt{x_{1} x_{2} \xi_1
M_{\Lambda_{b}} M_{\Lambda_{c}}}b_{2})  K_{0}( \sqrt{ x_{1} \xi_1
M_{\Lambda_{b}}
M_{\Lambda_{c}}}|\textbf{b}_1+\textbf{b}_2|),\,\,\,\,\,\,
\end{eqnarray}

with
\begin{eqnarray*}
&&t_a=max( \sqrt{ x_{2} \xi_1 M_{\Lambda_{b}}
M_{\Lambda_{c}}},1/|\textbf{b}_1+\textbf{b}_2|,1/b_{1}, 1/b_{2}),\\
&&t_b=max( \sqrt{ x_{1} \xi_1 M_{\Lambda_{b}}
M_{\Lambda_{c}}},1/|\textbf{b}_1+\textbf{b}_2|,1/b_{1}, 1/b_{2}).
\end{eqnarray*}


\end{document}